%% file: mass.tex
\documentclass[usenatbib]{mn2e}
\usepackage{amssymb,amsmath}
\include{epsf}

\def \farcs{\hbox{$.\!\!^{\prime\prime}$}}

\def \msun{\hbox{M$_\odot$}}

\title[LSS effects on weak lensing masses]{Effects of large-scale
  structure on the accuracy of weak lensing mass measurements}

\author[H. Hoekstra et al.]{Henk Hoekstra$^1$, Jan Hartlap$^2$, Stefan
  Hilbert$^{2,3}$ \& Edo van Uitert$^1$\\ 
$^1$Leiden Observatory, Leiden University, Niels Bohrweg 2, 2333 CA Leiden, The  Netherlands\\ 
$^2$Argelander-Institut f{\"u}r Astronomie, Universit{\"a}t Bonn, Auf dem H{\"u}gel 71, 53121 Bonn, Germany\\ 
$^3$Max Planck Institute for Astrophysics, Karl-Schwarzschild-Str. 1, 85741 Garching, Germany\\}

\begin{document}

\date{Accepted. Received; in original form}

\maketitle

\begin{abstract}

Weak gravitational lensing has become an important method to determine
the masses of galaxy clusters. The intrinsic shapes of the galaxies
are a dominant source of uncertainty, but there are other limitations
to the precision that can be achieved. In this paper we revisit a
typically ignored source of uncertainty: structure along the line-of
sight. Using results from the Millennium Simulation we confirm the
validity of analytical calculations that have shown that such random
projections are particularly important for studies of the cluster
density profile. In general the contribution of large-scale structure
to the total error budget is comparable to the statistical errors.  We
find that the precision of the mass measurement can be improved only
slightly by modelling the large-scale structure using readily available
data.

\end{abstract}


\section{Introduction}

The number density of galaxy clusters as a function of mass and
redshift can be used to constrain cosmological parameters and probe
the growth of structure
\citep[e.g.,][]{Henry00,Borgani01,Gladders07,Henry09,Mantz10,Vikhlinin09}.
This approach is conceptually straightforward, but the actual
implementation of this method is more difficult. This is because the
clusters are identified based on their baryonic properties (e.g.,
galaxy counts, SZ decrement, X-ray luminosity or temperature), which
need to be related to the underlying dark matter distribution.  The
relation between observed cluster properties and the mass depends
itself on the relative importance of the various physical processes
that play a role in galaxy and cluster formation. Galaxy clusters
provide an excellent laboratory to study these, because
multi-wavelength observations provide us with a complete census of the
various components: we can actually observe the stars, gas and dark
matter content.

The use of clusters to constrain cosmology and to improve our
understanding of cluster physics are closely inter-related: feedback
processes and differences in formation history lead to variations in
the observable properties at a given mass. The resulting intrinsic
scatter changes the selection function and thus leads to biased
constraints on cosmological parameters if left unaccounted for.
Correctly interpreting the scatter in the scaling relations requires a
good understanding of the various sources of uncertainty, which can be
either physical or statistical.

A variety of methods can be used to determine the mass of a cluster.
Most of these are based on dynamics and assume the cluster is relaxed.
In this case, the mass can obtained from the velocity dispersion of
the cluster galaxies. Measurements of the gas pressure, obtained from
observations of the hot X-ray emitting intracluster medium (ICM),
provide another powerful tracer of the dark matter content.  This
approach, which assumes hydrostatic equilibrium, has been used
extensively, thanks to the high quality observations obtained using
powerful X-ray telescopes such as the {\it Chandra} X-ray Observatory
and {\it XMM-Newton} \citep[e.g.,][]{Vikhlinin09,Mantz10}.
The interpretation of such measurements is often complicated by the
presence of substructures and the fact that most clusters are not relaxed.
A major concern is the assumption of hydrostatic equilibrium,
because numerical simulations have shown that active galactic nuclei,
turbulence, and bulk motions of the gas, as well as variations in the
merging history can lead to systematic underestimates of masses based
on X-ray observations \citep[e.g.,][]{Evrard90, Dolag05, Rasia06,
  Nagai07}. Although recent simulations incorporate a wide range of
physical processes it is not clear to what extent the simulations
provide a realistic estimate of the systematic error in the mass. It
is therefore critical to compare dynamical techniques to methods that
do not suffer from these problems.

This is possible thanks to a phenomenon called weak gravitational
lensing: the cluster mass distribution perturbs the paths of photons
emitted by distant galaxies. As a result the images of these
background galaxies appear slightly distorted (or sheared). Unlike
dynamical methods, gravitational lensing does not require one to make
assumptions regarding the dynamical state of the cluster. The
amplitude of this distortion provides us with a direct measurement of
the gravitational tidal field, which in turn provides us with an
estimate for the projected cluster mass. The recent improvements in
sample size and precision allowed \cite{Mahdavi08} to compare weak
lensing and hydrostatic masses for a sample of 18
clusters. \cite{Mahdavi08} found that at large radii the X-ray results
underestimate the mass, in agreement with the findings from numerical
simulations. These findings demonstrate the usefulness of weak lensing
for multi-wavelength studies of galaxy clusters.

Weak lensing masses are now routinely measured for large samples of
clusters \citep[e.g.][]{Hoekstra07,Bardeau07,Okabe10}. Tests on
simulated data have shown that the best analysis methods can reduce
systematic errors in the shear measurements to $\sim 1-2\%$
\citep[][]{STEP1,STEP2}. Much of the recent progress has been driven
by the desire to measure the weak lensing signal caused by intervening
large-scale structure, a.k.a. cosmic shear \citep[for a recent review
  see][]{HJ08}. The cosmic shear signal has now been detected at high
significance in a number of surveys \citep[e.g.,][]{Hoekstra02,
  Waerbeke05, Hoekstra06, Benjamin07, Fu08, Schrabback10} and is one
of the most promising tools to study dark energy. 

This cosmological signal, however, limits the accuracy with which
cluster masses can be determined: the observed lensing signal is a
combination of the cluster signal {\it and} cosmic shear. As first
discussed in \cite{Hoekstra01} the large-scale structure along the
line-of-sight is an additional source of noise, but does not bias the
measurement of the mass. As shown by \cite{Hoekstra03} this 'cosmic
noise' is particularly relevant when studying cluster mass density
profiles \citep[also see][]{Dodelson04}. Although the effects of
uncorrelated structures along the line-of-sight are often
acknowledged, their contribution to the formal error budget has
typically been ignored. This is somewhat surprising, given that there
is little doubt that cosmic shear has been measured.

In this paper we revisit the predictions presented in
\cite{Hoekstra01,Hoekstra03} using ray-tracing results from the
Millennium Simulation \citep{Springel05,Hilbert09}, demonstrating in
\S3 once more that cosmic noise should not be ignored in weak lensing
studies. We also quantify for the first time the noise introduced by
the finite sampling of the source redshift distribution. In \S4 we
examine whether cosmic noise can be suppressed using readily available
data.

\section{Cosmic noise}

The observed lensing signal is the combination of the shear induced by
the cluster mass distribution {\it and} that of other structures along
the line-of-sight. The expectation value of the latter vanishes, but
it does introduce additional variance in the cluster mass estimate.
The effect of this cosmic noise on weak lensing cluster studies can be
quantified analytically \citep{Hoekstra01,Hoekstra03} or using
numerical simulations \citep{White02}. Not surprisingly, these studies
have shown that the cosmic noise is most important when the cluster
signal itself becomes small: i.e., when data at large cluster-centric
radii are used, or when clusters at low redshifts are studied. Cosmic
noise, however, is also a considerable source of uncertainty for
clusters at intermediate redshifts.

Even for a massive cluster, the induced change in the shape of a
source galaxy's image is typically small compared to its intrinsic
ellipticity.  It is therefore convenient to azimuthally average the
tangential shear $\gamma_T$ and study its variation as a function of
radius. It can be related to the surface density through

\begin{equation}
\langle\gamma_T\rangle(r)=\frac{\bar\Sigma(<r) - \bar\Sigma(r)}
{\Sigma_{\rm crit}}=\bar\kappa(<r)-\bar\kappa(r),
\end{equation}

\noindent where $\bar\Sigma(<r)$ is the mean surface density within an
aperture of radius $r$, and $\bar\Sigma(r)$ is the mean surface
density on a circle of radius $r$. The convergence $\kappa$, or
dimensionless surface density, is the ratio of the surface density and
the critical surface density $\Sigma_{\rm crit}$, which is given by

\begin{equation}
\Sigma_{\rm crit}=\frac{c^2}{4\pi G}\frac{D_s}{D_l D_{ls}},
\end{equation}

\noindent where $D_l$ is the angular diameter distance to the
lens. $D_{s}$ and $D_{ls}$ are the angular diameter distances from the
observer to the source and from the lens to the source,
respectively. 

The variance in the azimuthally averaged tangential shear in an
annulus ranging from $r_1$ to $r_2$ caused by large-scale structure
along the line-of-sight is given by \citep{Hoekstra03}:

\begin{equation}
\sigma^2_{\rm LSS}(r_1,r_2)= 2\pi\int_0^\infty dl~l
P_\kappa(l) g^2(l,r_1,r_2), \label{eqmap}
\end{equation}

\noindent where the convergence power spectrum $P_\kappa(l)$ is given
by:

\begin{equation}
P_\kappa(l)=\frac{9 H_0^4 \Omega_m^2}{4 c^4}
\int_0^{w_H} dw \left(\frac{\bar W(w)}{a(w)}\right)^2 
P_\delta\left(\frac{l}{f_K(w)};w\right).
\end{equation}

\noindent Here $w$ is the radial coordinate, $a(w)$ the cosmic scale
factor, and $f_K(w)$ the comoving angular diameter
distance. $P_\delta(l;w)$ is the matter power spectrum. We consider
relatively small scales, and therefore need to account for the
non-linear evolution \citep[e.g.][]{Jain97, Schneider98,
  Hilbert09}. $\bar W(w)$ is the average ratio of angular
diameter distances $D_{ls}/D_{s}$ for a redshift distribution of
sources $p_w(w)$:

\begin{equation}
\bar W(w)=\int_w^{w_H} dw' p_w(w')\frac{f_K(w'-w)}{f_K(w')}.
\end{equation}

The function $g(l,r_1,r_2)$ in Eqn.~(\ref{eqmap}) is a filter of the
convergence power spectrum and is specified by our choice to consider
the azimuthally averaged tangential shear. We refer to
\cite{Hoekstra03} for a more detailed discussion, including the
expression for $g(l,r_1,r_2)$. In this paper we measure the
azimuthally averaged tangential shear as a function of radius
$r=(r_1+r_2)/2$, in bins that are $r_2-r_1=15$ arcseconds wide.

\subsection{Source redshift sampling}

In this paper we identify another source of error, which is important
for high redshift clusters and at small radii. It arises because the
amplitude of the lensing signal depends on the redshift of the
sources. At large distances from the cluster, the signal is obtained
by averaging over a relatively large number of galaxies, thus sampling
the average redshift distribution fairly well. However, at small radii
the number of sources is much smaller leading to a large variance in
the actual redshift distribution $n(z)$. This problem can be dealt
with using photometric redshifts for the sources, but the required
increase in observing time may make this difficult to achieve in
practice.  This sampling variance depends on the width of the redshift
distribution and can be estimated from observations of blank fields
\citep[e.g.,][]{Ilbert06,Ilbert09}. This estimate, however, does not
account for the clustering of the source galaxies, which increases the
scatter further. To include the effects of source clustering one can
place corresponding apertures in observed fields with redshifts and
measure the scatter. This approach, however, does require a rather large
survey area. To quantify how the {\it combination} of distant (i.e.,
uncorrelated) large-scale structure and variations in the source
redshift distribution affect weak lensing mass determinations we need
a realistic distribution of source galaxies, which themselves are part
of the large-scale structure. This requires mock data sets based on
cosmological numerical simulations of a large area.

\subsection{Numerical simulations}

We use results from the Millennium Simulation \citep{Springel05}, which is
a large $N$-body simulation following the evolution of $2160^3$ dark matter
particles in a periodic box with a side length of $500\,h^{-1}\,{\rm Mpc}$,
using a flat $\Lambda$CDM cosmology\footnote{The values
for the cosmological parameters that were adopted are: a matter density of
$\Omega_m=0.25$, a cosmological constant with $\Omega_\Lambda=0.75$, a Hubble
constant of $H_0=73$km/s/Mpc, a spectral index $n=1$ and a normalisation
$\sigma_8=0.9$ for the primordial power spectrum of density fluctuations.}.
The lensing signal is obtained from a careful ray-tracing analysis
presented in detail in \cite{Hilbert09}. The simulation is carried out
by dividing the periodically continued matter distribution of the
Millennium Simulation into $36$ slices out to $z=3$, each with a
thickness of $\approx 100\,h^{-1}\,{\rm Mpc}$. These are subsequently
projected onto lens planes, and a set of light rays is propagated
through the array of planes. Using a set of recursion relations, the
ray positions on each plane and the Jacobian matrices for the light
paths from the observer to each plane are obtained. Different
realizations are obtained by choosing different observer positions, in
our case yielding 512 patches of one square degree each. 
The periodic repetition of structures along the line-of-sight, which
is caused by the finite volume of the Millennium Simulation, is
minimized by choosing a line-of-sight direction that encloses
carefully chosen angles with the faces of the simulation box. The
advantage of this approach is that the matter distribution remains
continuous across slice boundaries, so that correlations on scales
larger than the thickness of the redshift slices are
maintained. 

\begin{figure}
\begin{center}
\leavevmode
\hbox{%
\epsfxsize=\hsize
\epsffile{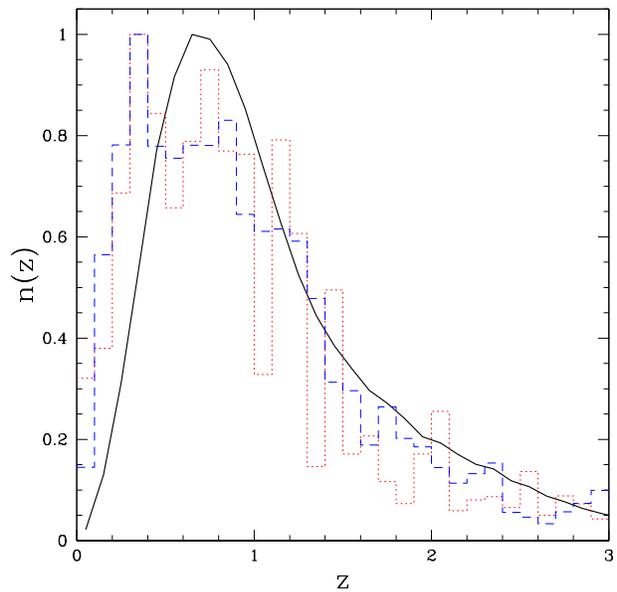}}
\begin{small}      
\caption{Redshift distribution $n(z)$ of the simulated source
  galaxies, with apparent magnitudes $r<25$ (solid line). For
  comparison redshift histograms for the COSMOS survey (red dotted
  histogram; Ilbert et al., 2009) and the CFHTLS (blue dashed
  histogram; Ilbert et al., 2005) are shown (same selection in
  apparent magnitude). The difference may be due to limitations of the
  simulation, but may also reflect incompleteness at high redshifts in
  the case of the photometric redshift catalogs.  Nonetheless, the
  redshift distribution derived from the simulations is adequate for
  our study.
\label{zdist}}
\end{small}
\end{center}
\end{figure}

Information on the properties of galaxies is obtained from
the semi-analytic models of galaxy formation by \cite{DeLucia07}.
Combined with the ray-tracing results, this allows us to obtain
realistic lensed positions and magnitudes for each galaxy, together
with shear and convergence at the galaxies' locations. For our galaxy
catalogues, we impose a magnitude cut of $r_{\rm SDSS}<25$.
The average redshift distribution of these source galaxies is presented
in Figure~\ref{zdist}. The results agree fairly well with photometric
redshift distributions from the COSMOS survey (\cite{Ilbert09}; red dotted
histogram) and the CFHT Legacy Survey (\cite{Ilbert06}; blue
dashed histogram). The actual redshift distributions appear to peak at
somewhat lower redshift, but this difference is not important for the
study presented here.

\subsection{Cluster Signal}

The numerical simulations provide a realistic lensing signal that
would be observed in a random patch of sky. This signal is also
present when a cluster of galaxies is studied: the observed lensing
signal is the combination of that of the cluster and the distant
large-scale structure (distant in the sense that it does not know
about the cluster). To simulate this, we can simply add the
cluster signal to that from the simulations. We assume that the
density profile of a cluster is described by the NFW \citep{NFW}
profile

\begin{equation}
\rho(r)=\frac{M_{\rm vir}}{4\pi f(c)}\frac{1}{r(r+r_s)^2},
\end{equation}

\noindent where $M_{\rm vir}$ is the virial mass, the mass enclosed
within the radius $r_{\rm vir}$. The virial radius is related to the
scale radius $r_s$ through the concentration $c=r_{\rm vir}/r_s$ and
the function $f(c)=\ln(1+c)-c/(1+c)$. By definition, the virial mass
and radius are related through

\begin{equation}
M_{\rm vir}=\frac{4\pi}{3} \Delta_{\rm vir}(z)\rho_{\rm bg}(z)r_{\rm vir}^3,
\end{equation}

\noindent where $\rho_{\rm bg}=3H_0^2\Omega_m(1+z)^3/(8\pi G)$ is the
mean density at the cluster redshift and the virial overdensity
$\Delta_{\rm vir}\approx (18\pi^2+82\xi-39\xi^2)/\Omega_m(z)$, with
$\xi=\Omega_m(z)-1$ \citep{Bryan98}. For the $\Lambda$CDM cosmology
considered here, $\Delta_{\rm vir}(0)=337$. Expressions for the
surface density and tangential shear of the NFW profile have been
derived by \cite{Bartelmann96} and \cite{Wright00} and we refer the
interested reader to these papers for the relevant equations.

In simulations of collisionless cold dark matter the NFW density
profile provides a good description of the radial mass distribution
for halos with a wide range in mass \citep[e.g.,][]{NFW95,NFW}. The
density profile is described by specifying $M_{\rm vir}$ and
concentration $c$ (or equivalently $r_s$). Numerical simulations,
however, indicate that the average concentration depends on the halo
mass and the redshift \citep{NFW95,Bullock01,Duffy08}. To account for
this correlation we use the relation between the virial mass $M_{\rm
  vir}$ and concentration $c$ from \cite{Duffy08} who studied
numerical simulations using the best fit parameters of the WMAP5
cosmology\footnote{This is a different cosmology from the one used to
  run the Millennium Simulation, but we note that the actual choice of
  mass-concentration relation is not important for the main results
  presented in this paper.} \citep{Komatsu09}. The best fit $c(M_{\rm
  vir})$ is given by:

\begin{equation}
c=7.85\left({\frac{M_{\rm vir}}{2\times 10^{12}}}\right)^{-0.081}{(1+z)^{-0.71}}.\label{cmrel}
\end{equation}

Simulations show considerable variation in the density profiles,
resulting in a lognormal distribution of $c$ with a scatter
$\sigma_{\log c}\sim 0.1$ for halos of a given mass
\citep[e.g.,][]{Jing00,Neto07}. Furthermore, \cite{Neto07} showed that
the concentration distributions are different for 'relaxed' and
'unrelaxed' halos. Although these physical variations are an
additional source of uncertainty when attempting to constrain the
mass-concentration relation observationally, they are not relevant for
our study of cosmic noise.

\section{Results}

\begin{figure}
\begin{center}
\leavevmode
\hbox{%
\epsfxsize=\hsize
\epsffile{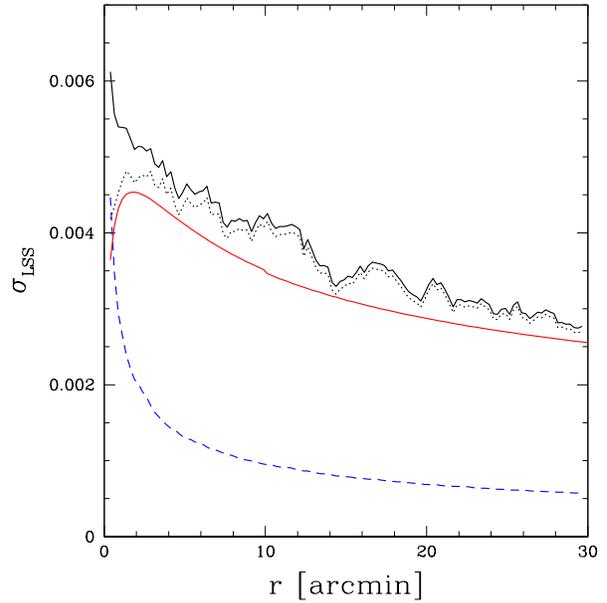}}
\begin{small}      
\caption{The dispersion $\sigma_{\rm
    LSS}=\langle\gamma_T^2\rangle^{1/2}$ introduced by distant large
  scale structure. The solid black line shows the dispersion measured
  from the 512 realisations from the Millennium Simulation.  At small
  radii the small number of sources introduces additional scatter
  (indicated by the dashed blue curve). The smooth red line
  corresponds to the analytical prediction from Hoekstra (2003). The
  prediction does not account for noise arising from the finite number
  of sources and should be compared to the dotted black line (which is
  corrected for this effect). The prediction is about 15\% lower,
  which is due to the adopted non-linear power spectrum (see
  text and Hilbert et al. 2009).
\label{gt_lss}}
\end{small}
\end{center}
\end{figure}

For each of the 512 realisations from the Millennium Simulation we
measure the azimuthally averaged tangential shear as a function of
radius from the centre of the image. The solid black line in
Figure~\ref{gt_lss} shows the resulting dispersion $\sigma_{\rm
  LSS}=\langle\gamma_T^2\rangle^{1/2}$ as determined from the 512
realisations. The prediction based on the \cite{PD96}
prescription for the non-linear evolution of the power spectrum is
indicated by the red line. On large scales ($>5'$) the prediction is
about 15\% lower, which we attribute to the adopted non-linear power
spectrum. This difference is consistent with the conclusions of
\cite{Hilbert09} who compared various prescriptions for the non-linear
power spectrum.

\begin{figure}
\begin{center}
\leavevmode
\hbox{%
\epsfxsize=\hsize
\epsffile{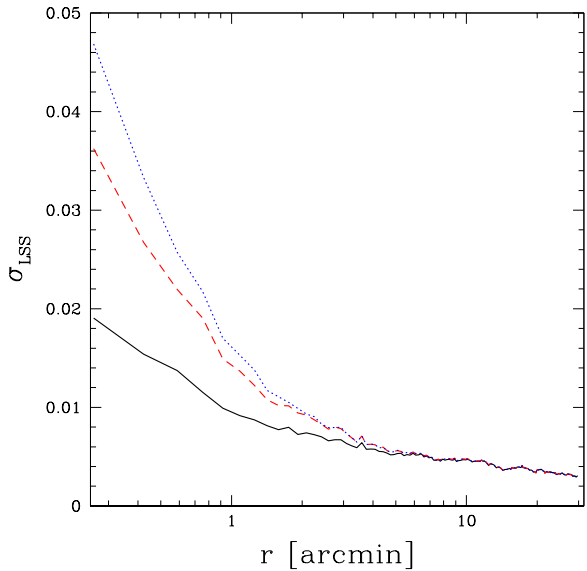}}
\begin{small}      
\caption{The dispersion $\langle \gamma_T^2\rangle^{\frac{1}{2}}$ for a cluster with $M_{\rm
    vir}=10^{15}M_\odot$ with redshift $z=0.2, 0.4$ and 0.6
  (solid-black, dashed-red and dotted blue lines, respectively). At large
  radii the cosmic shear contribution, which is independent of cluster
  redshift dominates the dispersion. On small scales, the dispersion
  in source redshifts increases with lens redshift.
\label{scat_all}}
\end{small}
\end{center}
\end{figure}

Close pairs of galaxies are sheared by similar amounts if all sources
are at the same redshift. In this case, the dispersion in the
tangential shear in a given radial bin would be small (for a given
realisation). This is not true for actual observations, because the
source redshift distribution is broad (see Figure~\ref{zdist}). At
large radii, where the signal is averaged over many galaxies, the
source redshift distribution is expected to be close to the average.
At small radii this is not a good representation, because the small
number of sources samples the average distribution only sparsely.
This leads to additional noise if photometric redshifts for the
sources are not available. Unlike the distant large-scale
structure, this effect is only relevant at small radii.

On small scales we can assume that the dispersion in the tangential
shear for a single realisation is predominantly caused by the spread
in source redshifts. The resulting mean dispersion as a function of
radius is indicated by the dashed blue curve in Figure~\ref{gt_lss}.
This is the contribution to $\sigma_{\rm LSS}$ caused by the finite
sampling of the source redshift distribution. If we remove this source
of scatter, the agreement between the \cite{Hoekstra03} prediction and
the Millennium Simulation is excellent (as indicated by the dotted
line), keeping in mind the difference in amplitude of the non-linear
power spectrum based on \cite{PD96}. Hence, variations in the actual
source redshift distribution lead to an increase in the observed
variance at radii $\lesssim 4'$. Most of the noise caused by the lack
of photometric redshifts arises from the fact the redshift
distribution is broad, but we expect that the scatter is boosted by
the fact that sources are in fact clustered. Comparison with the
simulations confirms this, but the increase in scatter is modest: the
increase is only $\sim 20\%$ compared to the estimate based on the
$n(z)$ alone.

The lack of knowledge of the actual source redshift distribution
contributes to the uncertainty in the cluster mass because it leads to
scatter in the ratio $\beta=D_{ls}/D_s$ in the expression for the
critical surface density.  For a low redshift cluster most sources are
at much higher redshifts and $\beta\sim 1$. Consequently the variation
in $\beta$ is small. As the redshift of the lens is closer to the mean
source redshift, the variation in $D_{ls}/D_s$ increases. This is
demonstrated in Figure~\ref{scat_all}, which shows the dispersion for
a cluster with mass $M_{\rm vir}=10^{15}M_\odot$ at various redshifts.
At large radii the variation in the redshift distribution is
negligible and the dispersion converges to the cosmic shear signal. At
small radii, the scatter caused by the variation in the source
redshift distribution increases rapidly with cluster redshift. Note
that deeper observations will improve the sampling of the redshift
distribution because of the larger number of sources. However, at the
same time the average source redshift will increase, resulting in a
larger cosmic noise contribution \citep[see][]{Hoekstra01}.

\subsection{Mass estimates}

In this section we study the combined effect of cosmic noise and the
finite sampling of the source redshift distribution on the weak
lensing mass estimate of a cluster with a virial mass $M_{\rm
  vir}=10^{15}M_\odot$. We assume it follows an NFW profile with the
concentration given by Eqn.~(\ref{cmrel}) and add the corresponding
lensing signal to the shear inferred from the ray-tracing analysis,
yielding 512 realisations of the cluster lensing signal.  We fit an
NFW model to the resulting lensing signal out to an outer radius
$R_{\rm out}$. The innermost point is $7\farcs5$, but the results do
not depend much on this choice: the small number of sources at these
radii means they have a low statistical weight in the fit. We consider
only shape noise as a source of uncertainty and determine the best fit
masses from a standard least-squares fit.

\begin{figure}
\begin{center}
\leavevmode
\hbox{%
\epsfxsize=\hsize
\epsffile{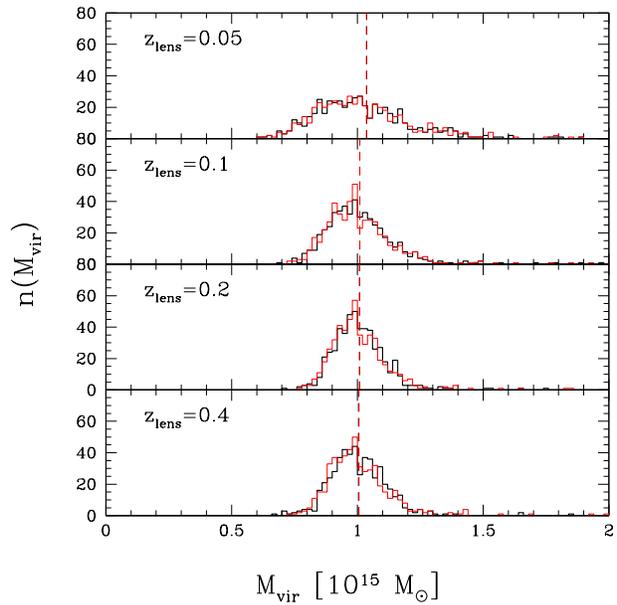}}
\begin{small}      
\caption{Histogram of the best fit virial masses $M_{\rm vir}$ for the
  512 realisations for lenses at $z=0.05,0.1,0.2$ and 0.4, when we
  adopt the mass-concentration relation from Duffy et al. (2008).  The
  red histograms show the distribution of results when only distant
  large scale structure is considered. For low redshifts the
  distribution of masses is skewed, but the average remains unbiased
  (indicated by the vertical dashed line). The black histograms show
  the results when the variation in the source redshift distribution
  is included. Note that these results do not include any shape noise.
\label{mvir_mc}}
\end{small}
\end{center}
\end{figure}

\begin{table}
\begin{center}
\caption{Dispersion in $M_{\rm vir}$ and $c$ \label{tabfree}}
\begin{tabular}{lcccccc}
\hline
\multicolumn{1}{c}{} &\multicolumn{2}{c}{adopting $c(M)$} &\multicolumn{4}{c}{$M_{\rm vir}$ \& $c$ free parameters} \\
\multicolumn{1}{c}{} &\multicolumn{1}{c}{$R_{\rm out}$=10'}&\multicolumn{1}{c}{$R_{\rm out}$=25'}&\multicolumn{2}{c}{$R_{\rm out}$=10'}&\multicolumn{2}{c}{$R_{\rm out}$=25'}\\
$z_{\rm lens}$ & $\sigma_M$ & $\sigma_M$ & $\sigma_M$ & $\sigma_c$ & $\sigma_M$ & $\sigma_c$ \\
\hline
\multicolumn{7}{c}{LSS only}\\
\hline
0.05 & 0.25 & 0.22 & 0.69 & 0.82 & 0.43 & 0.81 \\
0.1  & 0.15 & 0.17 & 0.35 & 0.53 & 0.25 & 0.63 \\
0.2  & 0.12 & 0.15 & 0.21 & 0.44 & 0.21 & 0.59 \\
0.4  & 0.13 & 0.18 & 0.21 & 0.48 & 0.23 & 0.69 \\
0.6  & 0.18 & 0.24 & 0.28 & 0.61 & 0.29 & 0.88 \\
\hline
\multicolumn{7}{c}{variation in $n(z)$ only}\\
\hline
0.05 & 0.01 & 0.01 & 0.01 & 0.04 & 0.01 & 0.03 \\
0.1  & 0.02 & 0.01 & 0.02 & 0.08 & 0.01 & 0.06 \\
0.2  & 0.03 & 0.02 & 0.03 & 0.25 & 0.02 & 0.23 \\
0.4  & 0.07 & 0.06 & 0.05 & 0.33 & 0.05 & 0.31 \\
0.6  & 0.10 & 0.08 & 0.08 & 0.45 & 0.07 & 0.41 \\
\hline
\multicolumn{7}{c}{combination of LSS and variation in $n(z)$}\\
\hline
0.05 & 0.25 & 0.22 & 0.73 & 0.83 & 0.44 & 0.81 \\
0.1  & 0.14 & 0.16 & 0.40 & 0.56 & 0.24 & 0.65 \\
0.2  & 0.11 & 0.14 & 0.23 & 0.54 & 0.21 & 0.67 \\
0.4  & 0.12 & 0.15 & 0.22 & 0.65 & 0.21 & 0.82 \\
0.6  & 0.16 & 0.22 & 0.28 & 0.85 & 0.26 & 1.08 \\
\hline
\multicolumn{7}{c}{statistical}\\
\hline
0.05 & 0.17 & 0.10 & 0.93 & 1.27 & 0.16 & 0.66 \\
0.1  & 0.12 & 0.08 & 0.25 & 0.81 & 0.10 & 0.54 \\
0.2  & 0.10 & 0.07 & 0.15 & 0.65 & 0.09 & 0.52 \\
0.4  & 0.12 & 0.10 & 0.15 & 0.73 & 0.11 & 0.63 \\
0.6  & 0.17 & 0.14 & 0.19 & 1.00 & 0.15 & 0.89 \\
\hline
\hline
\end{tabular}
\end{center}
{\footnotesize Dispersions in the values for the best-fit virial mass
  $M_{\rm vir}$ (in units of $10^{15}\msun$) and concentration $c$ as a
  function of lens redshift and the maximum radius that is used in the
  fit.  We list results for the effects of distant large-scale
  structure and variation in the source redshift distribution
  separately and combined. These results do not include the
  statistical error due to the intrinsic ellipticities of the source
  galaxies. The statistical errors (but now without LSS contributions)
  are given in the fourth set of results.}
\end{table}

We first consider the case where $M_{\rm vir}$ is the only parameter
that is fit, because the concentration is specified through
Eqn.~\ref{cmrel}. Figure~\ref{mvir_mc} shows the resulting
distribution of masses for clusters at various redshifts if we fit the
NFW model out to $R_{\rm out}=10'$. For low redshifts the distribution
is somewhat skewed, but the mean value is unbiased (as indicated by
the vertical dashed lines). Table~\ref{tabfree} lists the values for
the scatter $\sigma_M$ caused by the combined effects of cosmic noise
and source redshift variation. Comparison with the statistical errors
(computed assuming a total intrinsic source ellipticity of 0.3) shows
that the cosmic noise contribution is quite comparable at all
redshifts. Cosmic noise is minimal at intermediate redshifts
$(0.2<z<0.4)$.

\begin{figure}
\begin{center}
\leavevmode
\hbox{%
\epsfxsize=\hsize
\epsffile{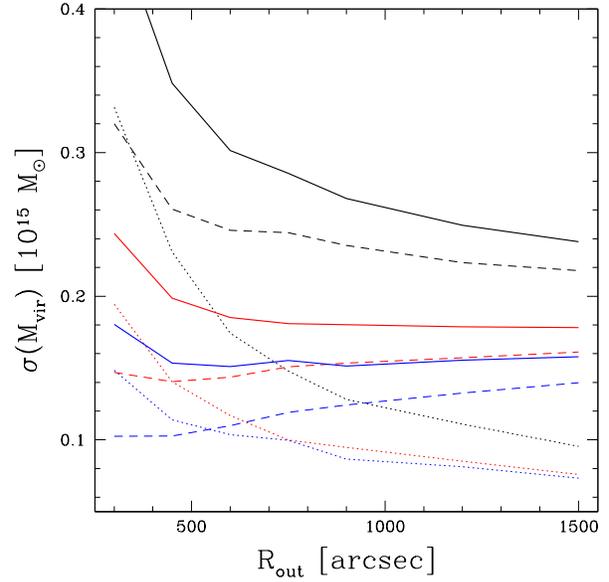}}
\begin{small}     
\caption{The solid lines show the total uncertainty in the best-fit
  virial mass (for $M_{\rm vir}=10^{15}\msun$) as a function of the
  maximum radius used to fit the NFW for a lens redshift of $z=0.05$
  (black), $z=0.1$ (red) and $z=0.2$ (blue). The contribution from the
  shape noise (i.e., statistical error) is indicated by the dotted
  curves, whereas the dashed lines show the LSS contribution. For
  $z=0.05$ there is a clear benefit from wide-field imaging data, but
  at $z=0.2$ the uncertainty is flat for $R_{\rm out}>10'$.
\label{scat_rout}}
\end{small}
\end{center}
\end{figure}

The increase for $z>0.4$ is caused by the fact that the angular extent
of the cluster decreases, whereas the aperture $R_{\rm out}$ is kept
fixed. The dashed lines in Figure~\ref{scat_rout} show how the cosmic
noise depends on $R_{\rm out}$. For reference, Table~\ref{tabfree}
also lists the values for $\sigma_M$ for $R_{out}=25'$. For $z=0.05$
(black line) the cosmic noise decreases with aperture size, but at
higher redshifts it increases. However, the statistical uncertainty
decreases with increasing $R_{\rm out}$, as indicated by the dotted
lines. The solid lines indicate the net result: for $z=0.05$ there is
a net gain, but for a cluster with $z>0.2$ (blue line) there is no
benefit extending the fit beyond $10'$.

We also compare the relative contributions of the distant large-scale
structure and the variation in $n(z)$. The latter is computed by using
the simulated redshift distribution, but without adding the cosmic
noise to the cluster signal. To compute the former, we add the cluster
shear computed for the average source redshift to the ray-tracing
results. Interestingly, the combined effect of LSS and variation in
$n(z)$ is to slightly reduce the scatter in the recovered masses,
compared to the LSS-only case. This can be easily understood: a
structure at lower redshift will increase the lensing signal, but will
also increase the number of sources at these redshifts. As the latter
are lensed less than the average source, they partly offset the
increase in lensing signal. Note that the combined effect does not
bias the cluster mass estimates.

The lensing signal of high redshift clusters can be boosted by
removing foreground galaxies using photometric redshift information
and optimally weighing the remaining sources based on their
$D_{ls}/D_s$. However, the cosmic noise increases also rapidly with
source redshift: $\sigma_{\rm LSS}\propto z^{1.4}$ for $(z<1)$, which
might limit the expected improvement in precision. The redshift
distribution used here drops quickly beyond $z\sim 1$ and we find that
for clusters with $z>0.4$ the photometric redshift information does
improve the mass measurements.

\subsection{Joint constraints on mass and concentration}

\begin{figure}
\begin{center}
\leavevmode
\hbox{%
\epsfxsize=\hsize
\epsffile{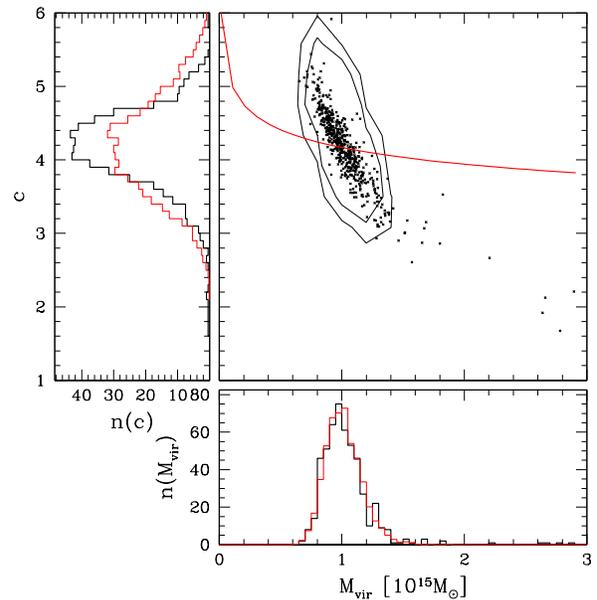}}
\begin{small}      
\caption{Distribution of best-fit $M_{\rm vir}$ and $c$ when both parameters
are free to vary. The points indicate the spread in results when distant
large-scale structure and source redshift variation are included. The contours
indicate the regions that enclose $68\%$ and $90\%$ of the fits when only
statistical (shape) noise is considered. The red line shows the mass-concentration
relation from Duffy et al. (2008). 
\label{mvirc}}
\end{small}
\end{center}
\end{figure}

So far we examined the effect of cosmic noise when one assumes a
mass-concentration relation which is based on numerical simulations.
Instead, many studies fit the lensing signal with both $M_{\rm vir}$
and $c$ as free parameters. This allows one to directly constrain the
concentration and therefore test the numerical simulations
\citep[e.g.,][]{Clowe02,Hoekstra02,Mandelbaum08,Okabe10}. Cosmic
noise, however, significantly increases the formal uncertainties in
such measurements \citep{Hoekstra03}. Figure~\ref{mvirc} shows the
distribution of best fit values for $M_{\rm vir}$ and concentration
$c$ for the 512 realisations when fitting both parameters
simultaneously. The contours indicate the statistical uncertainties in
the parameters, whereas the points show the spread due to cosmic noise
and variation in the source redshift distribution. Table~\ref{tabfree}
lists the scatter in the parameters. It is clear that cosmic noise has
a large impact on the ability to constrain the concentrations. In
particular, note the outliers with high inferred masses and low
concentrations. 

We also examined whether cosmic noise biases the slope of the
mass-concentration relation that is inferred from studies of samples
of clusters. For instance, \cite{Okabe10} obtain a power-law slope of
$0.40\pm0.19$, which is steeper than is seen in numerical
simulations. They examined the correlation in parameters due to the
shape noise for simulated profiles with masses $M_{\rm
  vir}=0.2-1.5\times 10^{15}\msun$. \cite{Okabe10} find a bias of
0.06, much smaller than the observed value. We performed a similar
test and find that cosmic noise also biases the inferred slope, but by
a similar amount.  The combined effect of shape and cosmic noise is a
bias of only 0.08 in the slope, because the correlation between
$M_{\rm vir}$ and $c$ is similar for both sources of error. We note
that the inferred slope is steeper if smaller range in mass is
considered: we find a bias of 0.17 for a range $M_{\rm
  vir}=1-1.5\times 10^{15}\msun$. Hence, the most important
consequence of including cosmic noise is to reduce the significance of the
measurement of the slope of the mass-concentration relation.

\section{Reducing cosmic noise}

The results presented in the previous section indicate that weak
lensing studies should include cosmic noise in their error budget. An
interesting question is whether one can reduce, or even
remove, the effects of cosmic noise. A statistical approach was
discussed by \cite{Dodelson04} who proposed a minimum variance
estimator to account for cosmic noise in mass reconstructions. A
concern, however, is that substructures associated with the cluster
might be suppressed as well. 

\subsection{Accounting for additional clusters}

In this section we will explore whether the observations themselves
can be used to reduce the cosmic noise. Although one can imagine many
different ways to predict the cosmic noise signal, we will consider a
relatively simple method. It requires only a minimum of colour
information and is therefore readily available: most studies include
(some) color information to identify cluster members.

Massive collapsed structures, such as galaxy clusters and groups of
galaxies contribute a large fraction of the power on the physical
scales relevant for cosmic noise. Fortunately they can be identified in
multi-colour data, similar to what is done in optical cluster
surveys. The most massive systems can readily be located using a
red-sequence method \citep[e.g.,][]{Gladders00}. Photometric
redshifts, involving more colours, can be used to find lower mass
halos. For instance, \cite{Milkeraitis10} used the Millennium
Simulation to examine how well one can identify clusters using five
optical filters. They find that clusters with masses larger than $\sim
5\times 10^{13}$ can be detected with fairly high completeness ($\sim
80\%$) and low false detection rate ($\sim 20\%$).

After having identified the clusters one needs to estimate their
contribution to the lensing signal. Here we take an optimistic
approach and assume we can find all halos down to a virial mass limit
$M_{\rm lim}$. In practice such a clear mass limit may be more
difficult to achieve. We fit these halos simultaneously with the
cluster of interest (where we ignore shape noise and assume the halos
follow an NFW profile with our adopted mass-concentration
relation). For a limiting mass $M_{\rm lim}=5\times 10^{13}\msun$ on
average 5.4 halos are fit in addition to the input cluster (with
actual numbers ranging from 0 to 12). We find that this procedure does
not bias the recovered cluster mass.

\begin{figure}
\begin{center}
\leavevmode
\hbox{%
\epsfxsize=\hsize
\epsffile{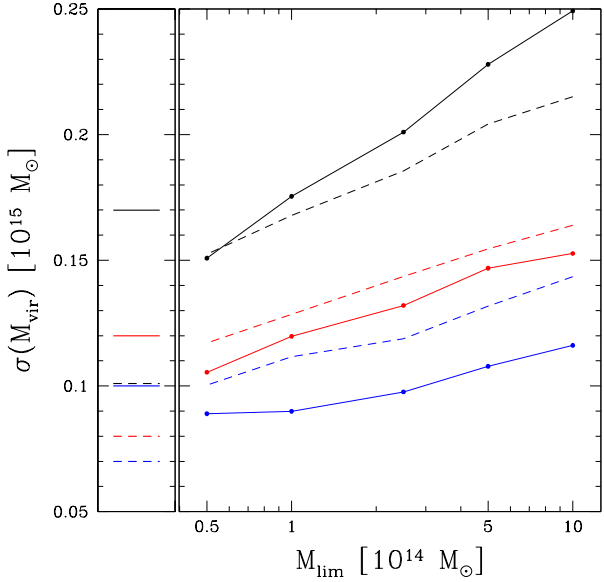}}
\begin{small}      
\caption{The scatter in the best-fit virial mass (for $M_{\rm
    vir}=10^{15}\msun$) as a function of $M_{\rm lim}$, the minimum
  mass of halos that are fit simultaneously with the cluster of
  interest. The improvement is largest for a cluster at $z=0.05$
  (black line). The benefits are smaller for $z=0.1$ (red line) and
  $z=0.2$ (blue line). The solid lines show the results when the NFW
  model is fit out to 10', whereas the dashed lines show the results
  for $R_{\rm out}=25'$. For reference, the left panel shows the
  statistical uncertainty in the virial mass due to the intrinsic
  shapes of the source galaxies.
\label{scat_mlim}}
\end{small}
\end{center}
\end{figure}

Figure~\ref{scat_mlim} shows the resulting scatter in the best fit
virial mass as a function of the mass limit of the halos included in
the fit. The solid lines show the results when the NFW model is fit
out to 10', whereas the dashed lines show the results for $R_{\rm
  out}=25'$. For reference, the left panel indicates the corresponding
statistical uncertainties in the virial mass due to the shape noise.

The results suggest our simple approach is indeed able to reduce the
effect of cosmic noise. Figure~\ref{scat_mlim} shows that this is most
relevant for clusters at very low redshifts. However, even with a
(low) mass limit of $M_{\rm lim}=5\times 10^{13}\msun$, the cosmic
noise remains a dominant source of uncertainty. We have examined several
other approaches, such as using the luminosities of galaxies to
predict the lensing signal, and found that none is able to
significantly improve upon the relatively simple approach outlined
above. We now will attempt to understand why this is.

\subsection{Limitations}

To compute the cosmic noise signal in \S3 we need the non-linear power
spectrum of density fluctuations to account for the fact that
collapsed halos increase the power on small scales. In general it is
computed using (fitting functions to) numerical simulations, such as
the prescription of \cite{PD96} that we used here. The observation
that dark matter halos are well described by NFW profiles allows for
an analytic approach as suggested by \cite{Seljak00}. In this model
the abundance of halos is given by the halo mass function and their
clustering is described by a mass dependent bias relation. The dark
matter profiles themselves are described by spherical NFW profiles
that are functions of the mass only (i.e., they follow a
mass-concentration relation).  The resulting power spectrum is the sum
of the contribution from a Poisson term that corresponds to individual
halos $P^{\rm P}(k)$ and a term arising from the clustering of halos
$P^{\rm hh}(k)$ themselves. This halo-model has proven to be useful to
study the clustering of galaxies and interpret the galaxy-mass
cross-correlation function. On small scales the Poisson term
dominates. It is given by

\begin{equation}
P^{\rm P}(k)=\frac{1}{(2\pi)^3}\int {\rm d}\nu f(\nu)\frac{M(\nu)}{\bar\rho}
\left|y(k,M(\nu))\right|^2,\label{ppoisson}
\end{equation}

\noindent where $\bar\rho$ is the mean matter density and $y(k,M)$ is
the ratio of the Fourier transform of the halo profile $\hat\rho(k)$
and the halo mass $M(\nu)$. The peak height $\nu$ of such an overdensity is
given by

\begin{equation}
\nu=\left[\frac{\delta_{\rm c}(z)}{\sigma(M)}\right]^2,
\end{equation}

\noindent where $\delta_{\rm c}$ is the value of a spherical
overdensity at which it collapses at a redshift $z$. $\sigma(M)$ is
the rms fluctuation in spheres that contain mass $M$ at an initial
time, extrapolated to $z$ using linear theory. The function $f(\nu)$
is related to the halo mass function ${\rm d}n/{\rm d}M$ through

\begin{equation}
\frac{{\rm d}n}{{\rm d}M}{\rm d}M=\frac{\bar\rho}{M}f(\nu){\rm d}\nu.
\end{equation}

\noindent We use the expressions from \cite{Sheth01} for $f(\nu)$ and the $M(c)$
relation from \cite{Duffy08} to compute the Poisson term. The halo-halo
term is important on large scales and is computed by integrating over the
mass function with the halo bias $b(\nu)$ and Fourier transform of the density
profile

\begin{equation}
P^{\rm hh}(k)=P_{\rm lin}(k)\left(\int {\rm d}\nu f(\nu) b(\nu) y(k,M(\nu))\right)^2,
\label{phh}
\end{equation}

\noindent where $P_{\rm lin}(k)$ is the linear power spectrum.

\begin{figure}
\begin{center}
\leavevmode
\hbox{%
\epsfxsize=\hsize
\epsffile{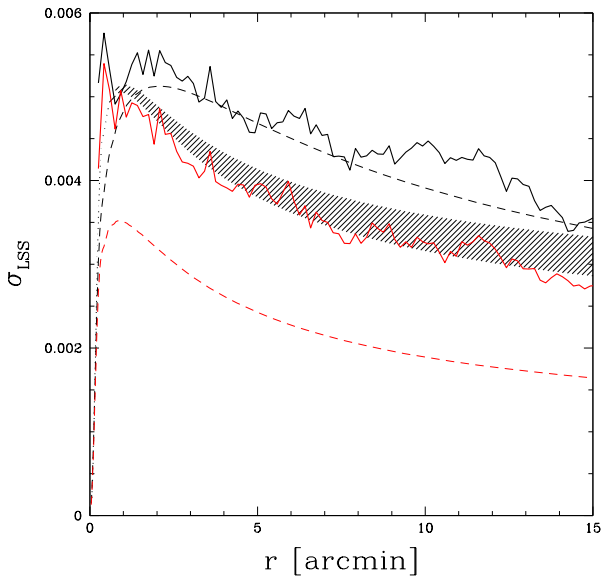}}
\begin{small}      
\caption{The cosmic noise signal (LSS-only) as a function of
  radius. The black line is the total signal and the dashed black line
  is the signal from the halo model. The Poisson term in the halo
  model has been multiplied by 1.15 to match the simulations. The
  solid red line indicates the cosmic noise when halos with $M_{\rm
    vir}>5\times 10^{13}\msun$ are included in the fit (see text for
  details). The red dashed line shows the halo model signal if such
  halos are removed perfectly. The dashed region indicates our
  estimate for the range in scatter when the uncertainties are taken
  into account.
\label{scat_cor}}
\end{small}
\end{center}
\end{figure}

The black line in Figure~\ref{scat_cor} shows the cosmic noise
(LSS-only) measured from the simulations. The dashed black line is the
halo model prediction, where the Poisson term has been multiplied by
1.15 to match the simulations. The solid red line in
Figure~\ref{scat_cor} shows $\sigma_{\rm LSS}$ if we fit all halos
with $M_{\rm vir}>5\times 10^{13}\msun$ as described in the previous
section. The theoretical limit of the reduction in power by accounting
for massive halos along the line-of-sight is obtained by integrating
Eqns.~(\ref{ppoisson}) and~(\ref{phh}) up to $M_{\rm lim}$, rather
than extending the integral over all masses. The dashed line in
Figure~\ref{scat_cor} shows the corresponding result for $M_{\rm
  lim}=5\times10^{13}\msun$. It is clear that this estimate
overestimates the reduction in cosmic noise, compared to the actual
simulated results.

The reason for this is simple: the theoretical limit implicitely
assumes that the halo masses were determined perfectly, which clearly
is too optimistic. Differences in the true mass $M_{\rm t}$ and the
fitted mass $M_{\rm f}$ add additional power to the theoretical limit.
For the Poisson term the residual power $P^{\rm P}{\rm res}$ is given
by:

\begin{eqnarray}
P^{\rm P}_{\rm res}(k)&=& \frac{1}{(2\pi)^3}\int
\limits^{\infty}_{M_{\rm lim}} {\rm d}\nu f(\nu)\frac{M_{\rm t}(\nu)}{\bar\rho}\times \nonumber\\
 & &\hspace{-1.0cm}\int\limits_0^\infty {\rm d}M_{\rm f}
 \left|y(k,M_{\rm t})-y(k,M_{\rm f})\right|^2W(M_{\rm t}-M_{\rm f}),
\end{eqnarray}

\noindent where $W(M_{\rm t}-M_{\rm f})$ describes the distribution of
the difference between the true and recovered masses. Comparison with
the simulations shows that $W$ can be approximated by a Gaussian with
a dispersion $\sigma$ that depends on the halo mass, with
$\sigma=2.3\times 10^{13}\msun+0.28\times M_{\rm t}$. We need to add
the contribution $P^{\rm hh}$ to $P^{\rm P}_{\rm res}$, which in the
ideal case is integrated up to $M_{\rm lim}$. However, we cannot fit
the contributions from halos outside the field of view and the actual
halo-halo contribution will lie between the ideal case and the full
$P^{\rm hh}$. 

The shaded region in Figure~\ref{scat_cor} indicates the expected
range in $\sigma_{\rm LSS}$ when we account for the uncertainties in
the modelling. The actual results agree very well with our estimates
based on the halo model. In reality the situation is even more dire,
because we ignored shape noise in the calculations presented here.

\section{Conclusions}

We used the Millennium Simulation to study how large-scale structure
along the line-of-sight (cosmic noise) affects the uncertainty in the
weak lensing masses of clusters of galaxies. After accounting for
differences in the calculation of the non-linear power spectrum of
density fluctuations, analytical estimates agree well with the
simulations. The simulations therefore support the findings by
\cite{Hoekstra01,Hoekstra03} that cosmic noise is a relevant source of
uncertainty in weak lensing mass determinations and therefore should
be included in the reported error budget. We do note that the adopted
$\sigma_8$ in the simulation is higher than the currently favoured
value, which reduces the amplitude of the cosmic noise somewhat.

We also examined whether variations in the source galaxy redshift
distribution are an important source of uncertainty. Although the
importance increases with the redshift of the cluster, we find it is
never significant when compared to statistical errors or cosmic noise.
For the simulated redshift distribution of sources used here we find that source
redshift information improves the precision of the mass measurement,
because the boost in lensing signal by the removal of foreground
galaxies is larger than the increase in cosmic noise due to the
increase in the mean source redshift.

Finally we examined whether it is possible to reduce the effect of
cosmic noise by identifying galaxy clusters and groups along the
line-of-sight. Such structures can be located fairly easily in
multi-colour data. We study a simple approach where we fit the masses
of these additional structures down to a mass $M_{\rm lim}$ and find
that cosmic noise can indeed be reduced, in particular for clusters at
very low redshifts $(z\sim 0.05)$. Nonetheless, the cosmic noise
remains a dominant source of uncertainty. To better understand the
limitations of modelling the contribution from distant large-scale
structure, we computed the expected signals using the halo model.  We
find that the uncertainties (or variations) in the profiles
fundamentally limit the suppression of cosmic noise. As a consequence,
cosmic noise will remain a dominant source of uncertainty in weak
lensing cluster mass measurements, and should not be ignored.

\section*{Acknowledgments} The authors would like to thank Tim Schrabback
for useful comments.  JH and SH acknowledge support by the Deutsche
Forschungsgemeinschaft within the Priority Programme 1177 under the
project SCHN 342/6 and by the German Federal Ministry of Education and
Research (BMBF) through the TR33 ``The Dark Universe''.  HH
acknowledges support from the Netherlands Organization for Scientific
Research (NWO) and a Marie Curie International Reintegration Grant.

\bibliographystyle{mn2e}
\bibliography{mass}

\end{document}



%% file: mass.bbl
\begin{thebibliography}{51}
\expandafter\ifx\csname natexlab\endcsname\relax\def\natexlab#1{#1}\fi

\bibitem[{{Bardeau} {et~al.}(2007){Bardeau}, {Soucail}, {Kneib}, {Czoske},
  {Ebeling}, {Hudelot}, {Smail}, \& {Smith}}]{Bardeau07}
{Bardeau} S., {Soucail} G., {Kneib} J., {Czoske} O., {Ebeling} H., {Hudelot}
  P., {Smail} I., {Smith} G.~P., 2007, \aap, 470, 449

\bibitem[{{Bartelmann}(1996)}]{Bartelmann96}
{Bartelmann} M., 1996, \aap, 313, 697

\bibitem[{{Benjamin} {et~al.}(2007){Benjamin}, {Heymans}, {Semboloni}, {van
  Waerbeke}, {Hoekstra}, {Erben}, {Gladders}, {Hetterscheidt}, {Mellier}, \&
  {Yee}}]{Benjamin07}
{Benjamin} J., {Heymans} C., {Semboloni} E., {van Waerbeke} L., {Hoekstra} H.,
  {Erben} T., {Gladders} M.~D., {Hetterscheidt} M., {Mellier} Y., {Yee}
  H.~K.~C., 2007, \mnras, 381, 702

\bibitem[{{Borgani} {et~al.}(2001){Borgani}, {Rosati}, {Tozzi}, {Stanford},
  {Eisenhardt}, {Lidman}, {Holden}, {Della Ceca}, {Norman}, \&
  {Squires}}]{Borgani01}
{Borgani} S., {Rosati} P., {Tozzi} P., {Stanford} S.~A., {Eisenhardt} P.~R.,
  {Lidman} C., {Holden} B., {Della Ceca} R., {Norman} C., {Squires} G., 2001,
  \apj, 561, 13

\bibitem[{{Bryan} \& {Norman}(1998)}]{Bryan98}
{Bryan} G.~L., {Norman} M.~L., 1998, \apj, 495, 80

\bibitem[{{Bullock} {et~al.}(2001){Bullock}, {Kolatt}, {Sigad}, {Somerville},
  {Kravtsov}, {Klypin}, {Primack}, \& {Dekel}}]{Bullock01}
{Bullock} J.~S., {Kolatt} T.~S., {Sigad} Y., {Somerville} R.~S., {Kravtsov}
  A.~V., {Klypin} A.~A., {Primack} J.~R., {Dekel} A., 2001, \mnras, 321, 559

\bibitem[{{Clowe} \& {Schneider}(2002)}]{Clowe02}
{Clowe} D., {Schneider} P., 2002, \aap, 395, 385

\bibitem[{{De Lucia} \& {Blaizot}(2007)}]{DeLucia07}
{De Lucia} G., {Blaizot} J., 2007, \mnras, 375, 2

\bibitem[{{Dodelson}(2004)}]{Dodelson04}
{Dodelson} S., 2004, \prd, 70, 023008

\bibitem[{{Dolag} {et~al.}(2005){Dolag}, {Vazza}, {Brunetti}, \&
  {Tormen}}]{Dolag05}
{Dolag} K., {Vazza} F., {Brunetti} G., {Tormen} G., 2005, \mnras, 364, 753

\bibitem[{{Duffy} {et~al.}(2008){Duffy}, {Schaye}, {Kay}, \& {Dalla
  Vecchia}}]{Duffy08}
{Duffy} A.~R., {Schaye} J., {Kay} S.~T., {Dalla Vecchia} C., 2008, \mnras, 390,
  L64

\bibitem[{{Evrard}(1990)}]{Evrard90}
{Evrard} A.~E., 1990, \apj, 363, 349

\bibitem[{{Fu} {et~al.}(2008){Fu}, {Semboloni}, {Hoekstra}, {Kilbinger}, {van
  Waerbeke}, {Tereno}, {Mellier}, {Heymans}, {Coupon}, {Benabed}, {Benjamin},
  {Bertin}, {Dor{\'e}}, {Hudson}, {Ilbert}, {Maoli}, {Marmo}, {McCracken}, \&
  {M{\'e}nard}}]{Fu08}
{Fu} L., {Semboloni} E., {Hoekstra} H., {Kilbinger} M., {van Waerbeke} L.,
  {Tereno} I., {Mellier} Y., {Heymans} C., {Coupon} J., {Benabed} K.,
  {Benjamin} J., {Bertin} E., {Dor{\'e}} O., {Hudson} M.~J., {Ilbert} O.,
  {Maoli} R., {Marmo} C., {McCracken} H.~J., {M{\'e}nard} B., 2008, \aap, 479,
  9

\bibitem[{{Gladders} \& {Yee}(2000)}]{Gladders00}
{Gladders} M.~D., {Yee} H.~K.~C., 2000, \aj, 120, 2148

\bibitem[{{Gladders} {et~al.}(2007){Gladders}, {Yee}, {Majumdar}, {Barrientos},
  {Hoekstra}, {Hall}, \& {Infante}}]{Gladders07}
{Gladders} M.~D., {Yee} H.~K.~C., {Majumdar} S., {Barrientos} L.~F., {Hoekstra}
  H., {Hall} P.~B., {Infante} L., 2007, \apj, 655, 128

\bibitem[{{Henry}(2000)}]{Henry00}
{Henry} J.~P., 2000, \apj, 534, 565

\bibitem[{{Henry} {et~al.}(2009){Henry}, {Evrard}, {Hoekstra}, {Babul}, \&
  {Mahdavi}}]{Henry09}
{Henry} J.~P., {Evrard} A.~E., {Hoekstra} H., {Babul} A., {Mahdavi} A., 2009,
  \apj, 691, 1307

\bibitem[{{Heymans} {et~al.}(2006){Heymans}, {Van Waerbeke}, {Bacon}, {Berge},
  {Bernstein}, {Bertin}, {Bridle}, {Brown}, {Clowe}, {Dahle}, {Erben}, {Gray},
  {Hetterscheidt}, {Hoekstra}, {Hudelot}, {Jarvis}, {Kuijken}, {Margoniner},
  {Massey}, {Mellier}, {Nakajima}, {Refregier}, {Rhodes}, {Schrabback}, \&
  {Wittman}}]{STEP1}
{Heymans} C., {Van Waerbeke} L., {Bacon} D., {Berge} J., {Bernstein} G.,
  {Bertin} E., {Bridle} S., {Brown} M.~L., {Clowe} D., {Dahle} H., {Erben} T.,
  {Gray} M., {Hetterscheidt} M., {Hoekstra} H., {Hudelot} P., {Jarvis} M.,
  {Kuijken} K., {Margoniner} V., {Massey} R., {Mellier} Y., {Nakajima} R.,
  {Refregier} A., {Rhodes} J., {Schrabback} T., {Wittman} D., 2006, \mnras,
  368, 1323

\bibitem[{{Hilbert} {et~al.}(2009){Hilbert}, {Hartlap}, {White}, \&
  {Schneider}}]{Hilbert09}
{Hilbert} S., {Hartlap} J., {White} S.~D.~M., {Schneider} P., 2009, \aap, 499,
  31

\bibitem[{{Hoekstra}(2001)}]{Hoekstra01}
{Hoekstra} H., 2001, \aap, 370, 743

\bibitem[{{Hoekstra}(2003)}]{Hoekstra03}
---, 2003, \mnras, 339, 1155

\bibitem[{{Hoekstra}(2007)}]{Hoekstra07}
---, 2007, \mnras, 379, 317

\bibitem[{{Hoekstra} \& {Jain}(2008)}]{HJ08}
{Hoekstra} H., {Jain} B., 2008, Annual Review of Nuclear and Particle Science,
  58, 99

\bibitem[{{Hoekstra} {et~al.}(2006){Hoekstra}, {Mellier}, {van Waerbeke},
  {Semboloni}, {Fu}, {Hudson}, {Parker}, {Tereno}, \& {Benabed}}]{Hoekstra06}
{Hoekstra} H., {Mellier} Y., {van Waerbeke} L., {Semboloni} E., {Fu} L.,
  {Hudson} M.~J., {Parker} L.~C., {Tereno} I., {Benabed} K., 2006, \apj, 647,
  116

\bibitem[{{Hoekstra} {et~al.}(2002){Hoekstra}, {Yee}, \&
  {Gladders}}]{Hoekstra02}
{Hoekstra} H., {Yee} H.~K.~C., {Gladders} M.~D., 2002, \apj, 577, 595

\bibitem[{{Ilbert} {et~al.}(2006){Ilbert}, {Arnouts}, {McCracken},
  {Bolzonella}, {Bertin}, {Le F{\`e}vre}, {Mellier}, {Zamorani}, {Pell{\`o}},
  {Iovino}, {Tresse}, {Le Brun}, {Bottini}, {Garilli}, {Maccagni}, {Picat},
  {Scaramella}, {Scodeggio}, {Vettolani}, {Zanichelli}, {Adami}, {Bardelli},
  {Cappi}, {Charlot}, {Ciliegi}, {Contini}, {Cucciati}, {Foucaud}, {Franzetti},
  {Gavignaud}, {Guzzo}, {Marano}, {Marinoni}, {Mazure}, {Meneux}, {Merighi},
  {Paltani}, {Pollo}, {Pozzetti}, {Radovich}, {Zucca}, {Bondi}, {Bongiorno},
  {Busarello}, {de La Torre}, {Gregorini}, {Lamareille}, {Mathez}, {Merluzzi},
  {Ripepi}, {Rizzo}, \& {Vergani}}]{Ilbert06}
{Ilbert} O., {Arnouts} S., {McCracken} H.~J., {Bolzonella} M., {Bertin} E., {Le
  F{\`e}vre} O., {Mellier} Y., {Zamorani} G., {Pell{\`o}} R., {Iovino} A.,
  {Tresse} L., {Le Brun} V., {Bottini} D., {Garilli} B., {Maccagni} D., {Picat}
  J.~P., {Scaramella} R., {Scodeggio} M., {Vettolani} G., {Zanichelli} A.,
  {Adami} C., {Bardelli} S., {Cappi} A., {Charlot} S., {Ciliegi} P., {Contini}
  T., {Cucciati} O., {Foucaud} S., {Franzetti} P., {Gavignaud} I., {Guzzo} L.,
  {Marano} B., {Marinoni} C., {Mazure} A., {Meneux} B., {Merighi} R., {Paltani}
  S., {Pollo} A., {Pozzetti} L., {Radovich} M., {Zucca} E., {Bondi} M.,
  {Bongiorno} A., {Busarello} G., {de La Torre} S., {Gregorini} L.,
  {Lamareille} F., {Mathez} G., {Merluzzi} P., {Ripepi} V., {Rizzo} D.,
  {Vergani} D., 2006, \aap, 457, 841

\bibitem[{{Ilbert} {et~al.}(2009){Ilbert}, {Capak}, {Salvato}, {Aussel},
  {McCracken}, {Sanders}, {Scoville}, {Kartaltepe}, {Arnouts}, {Le Floc'h},
  {Mobasher}, {Taniguchi}, {Lamareille}, {Leauthaud}, {Sasaki}, {Thompson},
  {Zamojski}, {Zamorani}, {Bardelli}, {Bolzonella}, {Bongiorno}, {Brusa},
  {Caputi}, {Carollo}, {Contini}, {Cook}, {Coppa}, {Cucciati}, {de la Torre},
  {de Ravel}, {Franzetti}, {Garilli}, {Hasinger}, {Iovino}, {Kampczyk},
  {Kneib}, {Knobel}, {Kovac}, {Le Borgne}, {Le Brun}, {F{\`e}vre}, {Lilly},
  {Looper}, {Maier}, {Mainieri}, {Mellier}, {Mignoli}, {Murayama}, {Pell{\`o}},
  {Peng}, {P{\'e}rez-Montero}, {Renzini}, {Ricciardelli}, {Schiminovich},
  {Scodeggio}, {Shioya}, {Silverman}, {Surace}, {Tanaka}, {Tasca}, {Tresse},
  {Vergani}, \& {Zucca}}]{Ilbert09}
{Ilbert} O., {Capak} P., {Salvato} M., {Aussel} H., {McCracken} H.~J.,
  {Sanders} D.~B., {Scoville} N., {Kartaltepe} J., {Arnouts} S., {Le Floc'h}
  E., {Mobasher} B., {Taniguchi} Y., {Lamareille} F., {Leauthaud} A., {Sasaki}
  S., {Thompson} D., {Zamojski} M., {Zamorani} G., {Bardelli} S., {Bolzonella}
  M., {Bongiorno} A., {Brusa} M., {Caputi} K.~I., {Carollo} C.~M., {Contini}
  T., {Cook} R., {Coppa} G., {Cucciati} O., {de la Torre} S., {de Ravel} L.,
  {Franzetti} P., {Garilli} B., {Hasinger} G., {Iovino} A., {Kampczyk} P.,
  {Kneib} J., {Knobel} C., {Kovac} K., {Le Borgne} J.~F., {Le Brun} V.,
  {F{\`e}vre} O.~L., {Lilly} S., {Looper} D., {Maier} C., {Mainieri} V.,
  {Mellier} Y., {Mignoli} M., {Murayama} T., {Pell{\`o}} R., {Peng} Y.,
  {P{\'e}rez-Montero} E., {Renzini} A., {Ricciardelli} E., {Schiminovich} D.,
  {Scodeggio} M., {Shioya} Y., {Silverman} J., {Surace} J., {Tanaka} M.,
  {Tasca} L., {Tresse} L., {Vergani} D., {Zucca} E., 2009, \apj, 690, 1236

\bibitem[{{Jain} \& {Seljak}(1997)}]{Jain97}
{Jain} B., {Seljak} U., 1997, \apj, 484, 560

\bibitem[{{Jing}(2000)}]{Jing00}
{Jing} Y.~P., 2000, \apj, 535, 30

\bibitem[{{Komatsu} {et~al.}(2009){Komatsu}, {Dunkley}, {Nolta}, {Bennett},
  {Gold}, {Hinshaw}, {Jarosik}, {Larson}, {Limon}, {Page}, {Spergel},
  {Halpern}, {Hill}, {Kogut}, {Meyer}, {Tucker}, {Weiland}, {Wollack}, \&
  {Wright}}]{Komatsu09}
{Komatsu} E., {Dunkley} J., {Nolta} M.~R., {Bennett} C.~L., {Gold} B.,
  {Hinshaw} G., {Jarosik} N., {Larson} D., {Limon} M., {Page} L., {Spergel}
  D.~N., {Halpern} M., {Hill} R.~S., {Kogut} A., {Meyer} S.~S., {Tucker} G.~S.,
  {Weiland} J.~L., {Wollack} E., {Wright} E.~L., 2009, \apjs, 180, 330

\bibitem[{{Mahdavi} {et~al.}(2008){Mahdavi}, {Hoekstra}, {Babul}, \&
  {Henry}}]{Mahdavi08}
{Mahdavi} A., {Hoekstra} H., {Babul} A., {Henry} J.~P., 2008, \mnras, 384, 1567

\bibitem[{{Mandelbaum} {et~al.}(2008){Mandelbaum}, {Seljak}, \&
  {Hirata}}]{Mandelbaum08}
{Mandelbaum} R., {Seljak} U., {Hirata} C.~M., 2008, JCAP, 8, 6

\bibitem[{{Mantz} {et~al.}(2010){Mantz}, {Allen}, {Rapetti}, \&
  {Ebeling}}]{Mantz10}
{Mantz} A., {Allen} S.~W., {Rapetti} D., {Ebeling} H., 2010, \mnras, 1029

\bibitem[{{Massey} {et~al.}(2007){Massey}, {Heymans}, {Berg{\'e}}, {Bernstein},
  {Bridle}, {Clowe}, {Dahle}, {Ellis}, {Erben}, {Hetterscheidt}, {High},
  {Hirata}, {Hoekstra}, {Hudelot}, {Jarvis}, {Johnston}, {Kuijken},
  {Margoniner}, {Mandelbaum}, {Mellier}, {Nakajima}, {Paulin-Henriksson},
  {Peeples}, {Roat}, {Refregier}, {Rhodes}, {Schrabback}, {Schirmer}, {Seljak},
  {Semboloni}, \& {van Waerbeke}}]{STEP2}
{Massey} R., {Heymans} C., {Berg{\'e}} J., {Bernstein} G., {Bridle} S., {Clowe}
  D., {Dahle} H., {Ellis} R., {Erben} T., {Hetterscheidt} M., {High} F.~W.,
  {Hirata} C., {Hoekstra} H., {Hudelot} P., {Jarvis} M., {Johnston} D.,
  {Kuijken} K., {Margoniner} V., {Mandelbaum} R., {Mellier} Y., {Nakajima} R.,
  {Paulin-Henriksson} S., {Peeples} M., {Roat} C., {Refregier} A., {Rhodes} J.,
  {Schrabback} T., {Schirmer} M., {Seljak} U., {Semboloni} E., {van Waerbeke}
  L., 2007, \mnras, 376, 13

\bibitem[{{Milkeraitis} {et~al.}(2010){Milkeraitis}, {van Waerbeke}, {Heymans},
  {Hildebrandt}, {Dietrich}, \& {Erben}}]{Milkeraitis10}
{Milkeraitis} M., {van Waerbeke} L., {Heymans} C., {Hildebrandt} H., {Dietrich}
  J.~P., {Erben} T., 2010, \mnras, 406, 673

\bibitem[{{Nagai} {et~al.}(2007){Nagai}, {Vikhlinin}, \& {Kravtsov}}]{Nagai07}
{Nagai} D., {Vikhlinin} A., {Kravtsov} A.~V., 2007, \apj, 655, 98

\bibitem[{{Navarro} {et~al.}(1995){Navarro}, {Frenk}, \& {White}}]{NFW95}
{Navarro} J.~F., {Frenk} C.~S., {White} S.~D.~M., 1995, \mnras, 275, 720

\bibitem[{{Navarro} {et~al.}(1997){Navarro}, {Frenk}, \& {White}}]{NFW}
---, 1997, \apj, 490, 493

\bibitem[{{Neto} {et~al.}(2007){Neto}, {Gao}, {Bett}, {Cole}, {Navarro},
  {Frenk}, {White}, {Springel}, \& {Jenkins}}]{Neto07}
{Neto} A.~F., {Gao} L., {Bett} P., {Cole} S., {Navarro} J.~F., {Frenk} C.~S.,
  {White} S.~D.~M., {Springel} V., {Jenkins} A., 2007, \mnras, 381, 1450

\bibitem[{{Okabe} {et~al.}(2010){Okabe}, {Takada}, {Umetsu}, {Futamase}, \&
  {Smith}}]{Okabe10}
{Okabe} N., {Takada} M., {Umetsu} K., {Futamase} T., {Smith} G.~P., 2010,
  \pasj, 62, 811

\bibitem[{{Peacock} \& {Dodds}(1996)}]{PD96}
{Peacock} J.~A., {Dodds} S.~J., 1996, \mnras, 280, L19

\bibitem[{{Rasia} {et~al.}(2006){Rasia}, {Ettori}, {Moscardini}, {Mazzotta},
  {Borgani}, {Dolag}, {Tormen}, {Cheng}, \& {Diaferio}}]{Rasia06}
{Rasia} E., {Ettori} S., {Moscardini} L., {Mazzotta} P., {Borgani} S., {Dolag}
  K., {Tormen} G., {Cheng} L.~M., {Diaferio} A., 2006, \mnras, 369, 2013

\bibitem[{{Schneider} {et~al.}(1998){Schneider}, {van Waerbeke}, {Jain}, \&
  {Kruse}}]{Schneider98}
{Schneider} P., {van Waerbeke} L., {Jain} B., {Kruse} G., 1998, \mnras, 296,
  873

\bibitem[{{Schrabback} {et~al.}(2010){Schrabback}, {Hartlap}, {Joachimi},
  {Kilbinger}, {Simon}, {Benabed}, {Brada{\v c}}, {Eifler}, {Erben},
  {Fassnacht}, {High}, {Hilbert}, {Hildebrandt}, {Hoekstra}, {Kuijken},
  {Marshall}, {Mellier}, {Morganson}, {Schneider}, {Semboloni}, {van Waerbeke},
  \& {Velander}}]{Schrabback10}
{Schrabback} T., {Hartlap} J., {Joachimi} B., {Kilbinger} M., {Simon} P.,
  {Benabed} K., {Brada{\v c}} M., {Eifler} T., {Erben} T., {Fassnacht} C.~D.,
  {High} F.~W., {Hilbert} S., {Hildebrandt} H., {Hoekstra} H., {Kuijken} K.,
  {Marshall} P.~J., {Mellier} Y., {Morganson} E., {Schneider} P., {Semboloni}
  E., {van Waerbeke} L., {Velander} M., 2010, \aap, 516, A63+

\bibitem[{{Seljak}(2000)}]{Seljak00}
{Seljak} U., 2000, \mnras, 318, 203

\bibitem[{{Sheth} {et~al.}(2001){Sheth}, {Mo}, \& {Tormen}}]{Sheth01}
{Sheth} R.~K., {Mo} H.~J., {Tormen} G., 2001, \mnras, 323, 1

\bibitem[{{Springel} {et~al.}(2005){Springel}, {White}, {Jenkins}, {Frenk},
  {Yoshida}, {Gao}, {Navarro}, {Thacker}, {Croton}, {Helly}, {Peacock}, {Cole},
  {Thomas}, {Couchman}, {Evrard}, {Colberg}, \& {Pearce}}]{Springel05}
{Springel} V., {White} S.~D.~M., {Jenkins} A., {Frenk} C.~S., {Yoshida} N.,
  {Gao} L., {Navarro} J., {Thacker} R., {Croton} D., {Helly} J., {Peacock}
  J.~A., {Cole} S., {Thomas} P., {Couchman} H., {Evrard} A., {Colberg} J.,
  {Pearce} F., 2005, \nat, 435, 629

\bibitem[{{Van Waerbeke} {et~al.}(2005){Van Waerbeke}, {Mellier}, \&
  {Hoekstra}}]{Waerbeke05}
{Van Waerbeke} L., {Mellier} Y., {Hoekstra} H., 2005, \aap, 429, 75

\bibitem[{{Vikhlinin} {et~al.}(2009){Vikhlinin}, {Kravtsov}, {Burenin},
  {Ebeling}, {Forman}, {Hornstrup}, {Jones}, {Murray}, {Nagai}, {Quintana}, \&
  {Voevodkin}}]{Vikhlinin09}
{Vikhlinin} A., {Kravtsov} A.~V., {Burenin} R.~A., {Ebeling} H., {Forman}
  W.~R., {Hornstrup} A., {Jones} C., {Murray} S.~S., {Nagai} D., {Quintana} H.,
  {Voevodkin} A., 2009, \apj, 692, 1060

\bibitem[{{White} {et~al.}(2002){White}, {van Waerbeke}, \& {Mackey}}]{White02}
{White} M., {van Waerbeke} L., {Mackey} J., 2002, \apj, 575, 640

\bibitem[{{Wright} \& {Brainerd}(2000)}]{Wright00}
{Wright} C.~O., {Brainerd} T.~G., 2000, \apj, 534, 34

\end{thebibliography}
